\documentclass[aps,prl,twocolumn,showpacs,preprintnumbers,amsmath,amssymb]{revtex4}

\usepackage{graphicx}
\usepackage{dcolumn}
\usepackage{color}
\usepackage{bm}
\usepackage{amsmath}

\usepackage{dcolumn}

\begin{document}

\title{Correlated Multiphoton Holes}
\author{I. Afek}
\author{O. Ambar}%
\author{Y. Silberberg}%

\affiliation{%
Department of Physics of Complex Systems, Weizmann Institute of
Science, Rehovot 76100, Israel }
\date{\today}
\begin{abstract}
We generate bipartite states of light which exhibit an \emph{absence} of multiphoton coincidence events between two modes amid a constant background flux. These `correlated photon holes' are produced by mixing a coherent state and relatively weak spontaneous parametric down-conversion using a balanced beamsplitter. Correlated holes with arbitrarily high photon numbers may be obtained by adjusting the relative phase and amplitude of the inputs. We measure states of up to five photons and verify their nonclassicality. The scheme provides a route for observation of high-photon-number nonclassical correlations without requiring intense quantum resources.
\end{abstract}
\maketitle

\begin{figure}[hhh!!]
\includegraphics{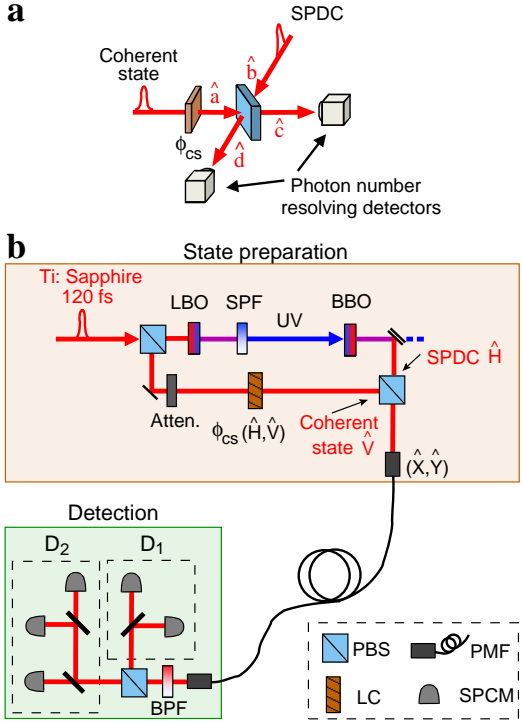}
\caption{\label{fig:setup} (Color online) Experimental setup for generation of correlated photon holes. \textbf{(a)} Schematic of the setup depicting a 50/50 beam-splitter fed by a coherent state and SPDC. The correlated  photon holes  occur in modes $\hat{c}$ and $\hat{d}$ after the beamsplitter by adjusting the relative phase and amplitude of the two input beams.  Measurement of multi-photon coincidence is performed using photon number resolving detectors. \textbf{(b)} Detailed layout of the setup. A pulsed Ti:Sapphire oscillator with $120$ fs FWHM pulse width and $80$ MHz repetition rate is doubled using a $2.74$mm LBO crystal to obtain $404$nm  ultra-violet pulses with maximum power of $225$mW. These pulses then pump collinear degenerate type-I SPDC at a wavelength of $808$nm  using a $1.78$mm long BBO crystal. The SPDC \mbox{($\hat{H}$ polarization)} is spatially and temporally overlapped with a coherent state ($\hat{V}$ polarization) using a polarizing beam-splitter cube (PBS). A thermally induced drift in the relative phase, $\phi_{cs}$, between $\hat{H}$ and $\hat{V}$ polarizations is corrected every few minutes using a liquid crystal (LC) phase retarder. The coherent light amplitude is adjusted using a variable attenuator. Modes $\hat{a}$($\hat{b}$) of panel (a) are realized using collinear $\hat{V}(\hat{H})$ polarizations respectively. These modes are then coherently mixed using  a polarization maintaining fiber (PMF) aligned with the $\pm45^{\circ}$ ($\hat{X},\hat{Y}$) polarization axes. The spatial and spectral modes are matched using the (single-mode) PMF and a $3$nm FWHM band-pass (BP) filter. Photon number resolving detection is performed using multiple single photon counting modules (SPCM, Perkin Elmer). Additional components: long-pass filter (LPF), short-pass filter (SPF).}
\end{figure}
\emph{Inroduction.}---
The generation of multiphoton entangled states has motivated a large body of experimental work in quantum optics \cite{PRL04EIS,PRL99BOUW,PRL01PAN}. The workhorse in such experiments has been spontaneous parametric down-conversion (SPDC), which allows generation of entangled photon pairs  \cite{PRL95KWIAT}. In recent years there has been an on going effort to create nonclassical states with more and more photons the highest value being a six photon graph state \cite{NATPHYS01PAN} using three SPDC pairs. Scaling up the number of photons in such schemes, however, is challenging since they rely on multiple emissions of SPDC pairs. This could be accomplished using state of the art, high intensity sources to pump the SPDC \cite{NATPHOT2010}.

Recently, the inverse of SPDC, namely the process in which a photon pair is \emph{missing} amid a constant background flux, has been demonstrated experimentally \cite{PRA06PIT}. This state has been dubbed an `entangled photon hole' and, just like regular entangled photons, it can be used to violate Bell's inequality \cite{EPHFRANSONPRL06}. Here we generalize this concept to more than two photons by creating two-mode states in which the probability for arbitrary photon numbers, $N_1,N_2$, to arrive simultaneously in the respective modes is zero, where choosing $N_1\!\!=N_2\!\!=1$  corresponds to the case of entangled photon holes \cite{EPHFRANSONPRL06,PRA06PIT}. We refer to the generated states as `correlated photon holes' (CPH). As in the two photon case \cite{PRA06PIT}, our scheme involves the mixing of SPDC and coherent light. Interestingly, the larger $N_1,N_2$ the higher the relative weight of the coherent light implying that our scheme may be implemented at high photon numbers with very modest SPDC fluxes. Boosting up the security of quantum cryptography with states similar to those generated here has been studied theoretically \cite{ADACHINJP2009,PRA2005LU}.

\emph{Theoretical scheme.}--- To date, a handful of photon counting experiments have utilized interference of coherent light and SPDC in a configuration sensitive to the relative phase. The SPDC in these was produced in either a single pass geometry \cite{PRA06PIT,NJP07PITTMAN,KOASHIPRL1993,AFEKSCIENCE} or an OPO \cite{PRL01OU}. With the exception of our recent demonstration of `high-NOON' states \cite{AFEKSCIENCE}, these experiments focused on two photon correlations. Here we generate another class of high-photon-number states which emerges naturally in this type of interference.

Consider a $50/50$ beam-splitter fed by a coherent state, $|\alpha\rangle_{a}$, in one input port and collinear degenerate SPDC, $|\xi\rangle_{b}$, in the other (see Fig. \ref{fig:setup}a). The input states are defined in the conventional way \cite{IntroQuantOPt}
\begin{eqnarray}
| \alpha \rangle &=& \sum_{n=0}^{\infty} e^{-\frac{1}{2}|\alpha|^2} \frac{\alpha^n}{\sqrt{n!}}|n\rangle, \quad \alpha = |\alpha|e^{\imath \phi_{cs}}
\notag\\
\ | \xi \rangle &=& \frac{1}{\sqrt{\cosh r }}\sum_{m=0}^{\infty}(-1)^m \frac{\sqrt{(2m)!}}{2^m m!} \notag \\
&{}& \quad \quad \quad \quad \quad \quad \quad \quad  \times(\tanh r)^m |2m\rangle,\label{eq:spdc_def}
\end{eqnarray}
where the phase of $|\xi\rangle$ has been set arbitrarily to zero leaving the relative phase of the two inputs to be determined by $\phi_{cs}$. We denote the pair amplitude ratio of the coherent state and SPDC  by
\begin{eqnarray}
\displaystyle \gamma \equiv |\alpha|^2 / r.\label{eq:gamma_def}
\end{eqnarray}
In physical terms $\gamma^2$ is the two photon probability of the classical source divided by that of the quantum source (in the limit $r,|\alpha|\ll1$). The larger the value of $\gamma$, the higher the relative weight of the classical resources. We denote the path entangled state at BS output modes $c,d$ (Fig. \ref{fig:setup}a) by $|\psi_{out}\rangle_{c,d}$. The amplitude for $N_1,N_2$ photons simultaneously in the BS output modes is then given by
\begin{eqnarray}
A_{N_1,N_2}=\langle N_1,N_2|\psi_{out}\rangle_{c,d}. \label{eq:defA}
\end{eqnarray}
In the absence of SPDC this is simply
\begin{eqnarray}
\displaystyle A_{N_1,N_2}=e^{-|\alpha/2|^2} (\alpha/2)^{N_1+N_2}/\sqrt{N_1! N_2!}, \label{eq:Anospdc}
\end{eqnarray}
which is non-zero for all values of $N_1,N_2$. By adding SPDC it is possible to cancel  $A_{N_1,N_2}$ for arbitrary values of $N_2,N_2$ by correctly adjusting $\gamma$ and $\phi_{cs}$. In the limit $r\ll1$, there are $\lfloor(N_1+N_2)/2\rfloor$ distinct solutions to the equation $A_{N_1,N_2}=0$ for any choice of $N_1,N_2$. For example for $N_1,N_2=1,1$ we choose $\gamma=1$ and $\phi_{cs}=\pi/2$. This produces an absence of $1,1$ coincidence events at the BS outputs which is similar to the previously studied case of two-photon holes \cite{PRA06PIT}. As a rule, the higher the values of $N_1,N_2$ the higher the value of $\gamma$, implying that relatively more photons originate from the classical source than from the SPDC. In the experimental part we demonstrate the scheme by measuring two cases, a four photon CPH with $N_1,N_2$=$2,2$ and $\gamma^2=3$ and a five photon CPH with $N_1,N_2$=$5,0$ and $\gamma^2=(15/(5-\sqrt{10}))^2\sim 66.58$. In general, the overall photon flux contributed by the SPDC is negligible compared to the coherent state flux. The SPDC may therefore be viewed as a small perturbation to the coherent field. It is noteworthy that our scheme bears some resemblance to a theoretical proposition for observation of antibunching using a degenerate parametric amplifier \cite{PRL74STOLER}.

\begin{figure}[t!!]
\includegraphics{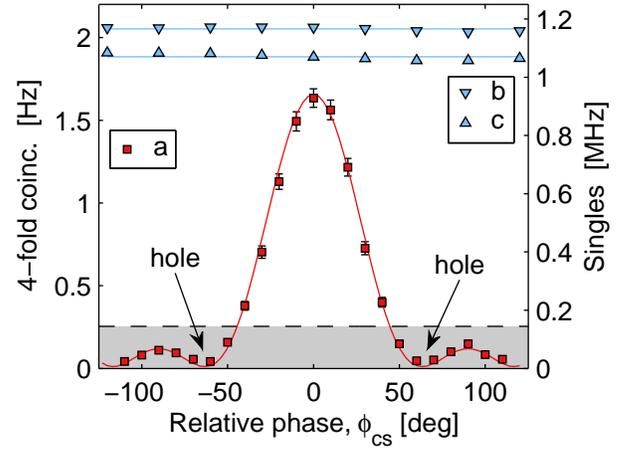}
\caption{\label{fig:22} (Color online) Measurement of a $2,2$ correlated photon hole. \textbf{(a)} (left y-axis) Coincidence events with two photons in $D_1$ and two photons in $D_2$ as a function of relative phase, $\phi_{cs}$, between the coherent state and the SPDC  (see Fig. \ref{fig:setup}). The minima correspond to phases in which the $2,2$ coincidence is canceled implying a correlated photon hole. The solid line is a theoretical calculation taking into account the overall setup transmission and detector positive-operator-valued-measures \cite{AFEKSCIENCE}. The visibility is $94.9\%$.  Error bars indicate $\pm \sigma$ statistical uncertainty. Dashed line indicates the classical bound for this measurement, see Eq. (\ref{eq:non-classical2}) and the preceding discussion. All points below this line (shaded area) exhibit non-classical behavior. The two arrows indicate the position of the correlated photon holes. \textbf{(b)},\textbf{(c)} (right y-axis) The single counts in detectors $D_1, D_2$ as a function of $\phi_{cs}$ which exhibit virtually no phase dependence, straight solid line is a guide to the eye.
}
\end{figure}

\begin{figure}[ht!]
\includegraphics{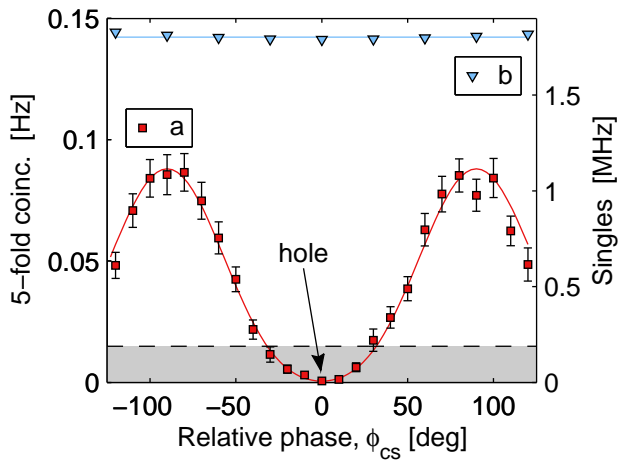}
\caption{\label{fig:50} (Color online) Measurement of a $5,0$ correlated photon hole. Here, all five detectors were employed in $D_1$ by placing a $\hat{H}$ transmitting polarizer followed by a half-wave-plate just before the PBS in the detector box. This allows us to measure five photon events in $D_1$ (see Fig. \ref{fig:setup}). \textbf{(a)} (left y-axis) Coincidence events with five photons in $D_1$ as a function of relative phase, $\phi_{cs}$, between the coherent state and the SPDC. The minimum, indicated by an arrow, corresponds to the phase in which the $N_1=5, N_2=0$ coincidence is canceled, implying a correlated five-photon hole. The solid line is a theoretical calculation taking into account the overall setup transmission and detector positive-operator-valued-measures \cite{AFEKSCIENCE}. The visibility is $98.6\%$. Error bars indicate $\pm \sigma$ statistical uncertainty. Dashed line indicates the classical bound for this measurement, see Eq. (\ref{eq:non-classical2}) and the preceding discussion. All points below this line (shaded area) exhibit non-classical behavior.  \textbf{(b)} (right y-axis) The single counts in detector $D_1$ as a function of $\phi_{cs}$ which exhibit virtually no phase dependence, straight solid line is a guide to the eye.
}
\end{figure}

\emph{Experimental setup and results.}---Our setup (Fig. \ref{fig:setup}) is similar to the one we have used for generation of NOON states with high photon numbers in a recent experiment \cite{AFEKSCIENCE}. The scheme requires generation of SPDC and coherent light with common spatial and spectral modes. The beams are prepared in perpendicular linear polarizations $(\hat{H},\hat{V})$ and overlapped using a polarizing beam-splitter cube (PBS). The phase between the beams, $\phi_{cs}$, is controlled before the PBS using a liquid crystal (LC) phase retarder. The BS of Fig. 1a is then implemented in a collinear geometry using a polarization maintaining fiber with axes aligned at $\pm 45^{\circ}$ $(\hat{X},\hat{Y})$.  Multiphoton coincidences with $N_1,N_2$ photons arriving simultaneously in detectors $D_1,D_2$ respectfully are measured as a function of $\phi_{cs}$ using an array of avalanche photodiodes. We apply the scheme for two pairs of coincident photon numbers. First, we create a four photon hole by choosing $N_1,N_2=2,2$ (Fig. \ref{fig:22}) i.e. the simultaneous arrival of two photons in  $D_1$ and two photons in $D_2$ is canceled while keeping a constant background flux. For this case $\gamma^2=3$ (see Eq. (\ref{eq:gamma_def})).  Next, we create a five photon hole by choosing $N_1,N_2=5,0$ (Fig. \ref{fig:50}) i.e. we cancel the probability for simultaneous arrival of five photons in $D_1$ and zero in $D_2$. In practice, for this case, we used all five of our single photon detectors (Perkin Elmer SPCM's) in $D_1$ and therefore could not post select events with zero counts in $D_2$. At the employed flux, however, the higher order $5,1$ coincidence events (which could have been eliminated by post selection) are extremely rare with only a minor effect on the measured signal and may be neglected. For the five photon case,  $\gamma^2=66.58$, implying that the coherent state two photon probability is $66.58$ times larger than that of the SPDC. Remarkably, such a weak perturbation of quantum light has a dramatic effect on the measured five photon events. The single counts, on the other hand, show no phase dependence in either case as seen in Figs. \ref{fig:22} and \ref{fig:50} (triangles). This is because the SPDC process generates only even photon numbers with no single photon events to interfere with those of the coherent state.

Due to losses, higher order events have the effect of adding a background to the $N$ photon coincidence measurement. This reduces the experimental visibilities and is the main limiting factor in observing photon holes with higher photon numbers. The overall transmission in our setup, which accounts for detector efficiencies and other sources of loss, is $12.5 \%$. This is measured by taking the ratio of coincidence to singe counts using only the SPDC beam (coherent state blocked). The solid lines accompanying the experimental plots (Figs. 2 and 3) are calculated using an analytical model of the experiment taking into account the transmission and the detector positive-operator-valued-measures \cite{AFEKSCIENCE}.

\emph{Nonclassical properties.}---
To quantify the nonclassicality of the generated states we derive a classical bound for Glauber's $nth$ order, equal-time, correlation function  $g^{(n)}(\tau\!\!=\!\!0)$. Note that we chose $\tau=0$ since we are interested in photons arriving to the detectors at the same time. This correlation function is proportional to the $n$ photon coincidence signal of a single spatial mode divided by the single count raised to the $nth$ power.  For thermal light, which exhibits bunching, $g^{(n)}(0)=n!$ and for a coherent state $g^{(n)}(0)=1$ \cite{IntroQuantOPt,OEXPRESS2010STEVENS}. In the following we show that the result for a coherent state is actually the lower bound for an arbitrary classical state i.e. a state with an arbitrary nonnegative well behaved $P$ function \cite{PhysRev65Glauber}. The inequality was implied in Glauber's work on high-order correlation functions of coherent fields  \cite{PhysRev65Glauber} and is derived here somewhat differently,
\begin{eqnarray}
\displaystyle
\displaystyle g^{(n)}(\tau \! = \! 0)=\frac{ \mbox{Tr} \left\{ \hat{\rho}\hat{E}^{(-)^n} \hat{E}^{(+)^n}  \right\} }
{\mbox{Tr} \left\{ \hat{\rho}\hat{E}^{(-)} \hat{E}^{(+)} \right\}^n} \notag
\\ = \frac{\int P(\alpha) |\alpha|^{2n}d^2\alpha}{\left(\int P(\alpha) |\alpha|^2 d^2\alpha\right)^n} \geq 1,\label{eq:non-classical1}
\end{eqnarray}
where the last inequality follows immediately from a multidimensional form of Jensen's inequality \cite{note,JENSEN2009}. We note that for $n=2$ this inequality is a well known result of the Schwarz inequality \cite{IntroQuantOPt}. Correlated photon holes require a bound for the \emph{two-point} equal-time Glauber correlation function $g^{(m,n)}(x_1,x_2; \tau \! = \! 0)$, where $x_1,x_2$ correspond to the BS output modes (Fig 1a). The derivation is based on the single mode result, Eq. (\ref{eq:non-classical1}), and requires the assumption that the two mode $P$ function describing the BS outputs is separable i.e $P_{1,2}(\alpha,\beta)=P_1(\alpha) P_2(\beta)$, a condition which is satisfied by all classical two mode gaussian states \cite{LEEPRA2000}. Using this assumption the inequality follows immediately from the single mode result applied to each of the modes independently,
\begin{widetext}
\begin{eqnarray}
g^{(m,n)}(x_1,x_2; \tau \! = \! 0)&=&\frac{ \mbox{Tr}
\left\{ \hat{\rho}\hat{E}^{(-)^m}(x_1) \hat{E}^{(-)^n}(x_2) \hat{E}^{(+)^n}(x_2) \hat{E}^{(+)^m}(x_1)  \right\} }
{\mbox{Tr} \left\{ \hat{\rho}\hat{E}^{(-)}(x_1) \hat{E}^{(+)}(x_1)\right\}^m \mbox{Tr} \left\{\hat{\rho}\hat{E}^{(-)}(x_2) \hat{E}^{(+)}(x_2) \right\}^n} \notag
\\ &=& \frac{\int P_{1,2}(\alpha,\beta) |\alpha|^{2m} |\beta|^{2n}d^2\alpha d^2\beta}
{\left(\int P_1(\alpha) |\alpha|^2 d^2\alpha \right)^m \left(\int P_2(\beta) |\beta|^2 d^2\beta \right)^n}
= \frac{\int P_{1}(\alpha) |\alpha|^{2m}d^2\alpha}{\left(\int P_1(\alpha) |\alpha|^2 d^2\alpha \right)^m}
\frac{\int P_{2}(\beta) |\beta|^{2n}d^2\beta}{\left(\int P_2(\beta) |\beta|^2 d^2\beta \right)^n}  \geq 1. \label{eq:non-classical2}
\end{eqnarray}
\end{widetext}
We used Eq. (\ref{eq:non-classical2}) to calculate the shaded nonclassical areas in Figs. 2 and 3. For the four photon case (Fig. 2) we substitute $m=2,n=2$. The classical bound is violated by $23.92$ and $21.17$ standard deviations at the phases for which the photon holes are created. For the five photon case (Fig. \ref{fig:50}) we substitute $m=5,n=0$. The bound is violated by $25.5$ standard deviations at the five photon hole.

\emph{Conclusion.}--- Mixing coherent light and SPDC is typically done in conjunction with homodyne detection \cite{IntroQuantOPt} using a `macroscopic' local oscillator as the  coherent state. Adopting this paradigm in the `few photon' regime and using number resolving detectors brings to light a rich structure exhibiting various nonclassical signatures. This has enabled us to create NOON states \cite{AFEKSCIENCE} in a recent work and correlated photon holes here. Extending the present work to higher photon numbers can be done using essentially the same setup, requiring only additional detectors to enable higher coincidence measurements. This does not entail a larger SPDC flux since relatively more of the photons originate from the coherent state which is practically unlimited in intensity, providing experimental simplification. As in the case of NOON states however, the visibility of interference is eventually limited by the overall setup transmission (currently $12.5\%$), determined by accounting for all sources of photon loss including the detector efficiency . The transmission could be improved by using high purity SPDC sources which can be spectrally mode matched to coherent states without requiring a bandpass filter \cite{PRL08Lundeen,NJPHYS08WALMSLEY,PRL07Torres}. Improved single mode fiber coupling of the photon pairs and high efficiency photon number resolving detectors \cite{CP07,OE08} would also be beneficial. Reaching a transmission of $25\%$ would allow measuring states with ten photons in a setup similar to ours.

\begin{acknowledgments}
I.A. gratefully acknowledges the support of the Ilan Ramon fellowship.
Financial support of this research by the German Israel Foundation (GIF),
the Minerva Foundation and the Crown Photonics Center is gratefully acknowledged.
Correspondence and requests for materials
should be addressed to Afek. I ~(email: itai.afek@weizmann.ac.il).
\end{acknowledgments}


\begin{thebibliography}{12}
\bibitem{note} Jensen's inequality states that \\
$\displaystyle \mu \left( \int^\infty_\infty g(\mathbf{x})P(\mathbf{x})d\mathbf{x} \right) \leq\int_\infty^\infty \mu(g(\mathbf{x}))P(\mathbf{x})d\mathbf{x}$, \\
where $P$ is a probability density function, $g$ is a real valued measurable function and $\mu$ is a real convex function.
\end{thebibliography}
\end{document}